1  The distribution of interplanetary dust between 0.96 and 1.04 AU as inferred
2  from impacts on the STEREO spacecraft observed by the Heliospheric Imagers

4  C. J. Davis[1,2,*], J. A. Davies[1], O. C St Cyr[3], M. Campbell-Brown[4], A. Skelt[1,5],
5  M. Kaiser[3], Nicole Meyer-Vernet[6], S. Crothers[1], C. Lintott[7,8], A. Smith[7,8], S. Bamford[9],
6  E. M. L. Baeten[10].

7  [1]RAL Space, Rutherford Appleton Laboratory, Chilton, Oxfordshire, UK.
8  [2]Department of Meteorology, University of Reading, Berkshire, UK.
9  [3] Goddard Space Flight Centre, USA.
10  [4] University of Western Ontario, London, Canada.
11  [5] Faringdon Community College, Farringdon, Oxfordshire, UK.
12  [6]Observatoire de Paris, Meudon Cedex, France.
13  [7]Astrophysics Department, University of Oxford, UK.
14  [8]Adler Planetarium, Chicago, USA.
15  [9]Centre for Astronomy and Particle Theory, The University of Nottingham, University Park, Nottingham, NG7
16  2RD, UK
17  [10]Zooniverse, c/o Astrophysics Department, University of Oxford, UK.

19  Corresponding author: C. J. Davis (chris.davis@stfc.ac.uk)

20  **ABSTRACT**

21  The distribution of dust in the ecliptic plane between 0.96 and 1.04 AU has been inferred from
22  impacts on the two STEREO spacecraft through observation of secondary particle trails and
23  unexpected off-points in the Heliospheric Imager (HI) cameras. This study made use of analysis
24  carried out by members of a distributed web-based citizen science project, Solar Stormwatch. A
25  comparison between observations of the brightest particle trails and a survey of fainter trails shows
26  consistent distributions.
27  While there is no obvious correlation between this distribution and the occurrence of individual
28  meteor streams at Earth, there are some broad longitudinal features in these distributions that are
29  also observed in sources of the sporadic meteor population. The different position of the HI
30  instrument on the two STEREO spacecraft leads to each sampling different populations of dust
31  particles. The asymmetry in the number of trails seen by each spacecraft and the fact that there are
32  many more unexpected off-points in the HI-B than in HI-A, indicates that the majority of impacts are
33  coming from the apex direction. For impacts causing off-points in the HI-B camera these dust
34  particles are estimated to have masses in excess of $10^{-17}$ kg with radii exceeding 0.1 µm.

For off-points observed in the HI-A images, which can only have been caused by particles travelling from the anti-apex direction, the distribution is consistent with that of secondary 'storm' trails observed by HI-B, providing evidence that these trails also result from impacts with primary particles from an anti-apex source. Investigating the mass distribution for the off-points of both HI-A and HI-B, it is apparent that the differential mass index of particles from the apex direction (causing off-points in HI-B) is consistently above 2. This indicates that the majority of the mass is within the smaller particles of this population. In contrast, the differential mass index of particles from the anti-apex direction (causing off-points in HI-A) is consistently below 2, indicating that the majority of the mass is to be found in larger particles of this distribution.

**1. INTRODUCTION**

The twin spacecraft of the NASA STEREO mission (Russell et al, 2008) were designed to enable studies of the Sun and inner heliosphere in three dimensions. The spacecraft are in heliocentric ecliptic orbits similar to the Earth, one drifting ahead of the Earth (STEREO-Ahead or STEREO-A) at a distance of approximately 0.96 Astronomical Units (AU) from the Sun while the other (STEREO-Behind or STEREO-B) lags behind the Earth at a distance of approximately 1.04 AU from the Sun. As a consequence of their orbits, the two observatories are moving away from the Earth at a rate that causes the Earth-Sun-spacecraft angle to increase by around $22.5^o$ per year. Each spacecraft carries a suite of instrumentation to image the Sun and solar wind and make in-situ measurements of the interplanetary medium.

Since their launch in 2006, these spacecraft have been detecting impacts by interplanetary dust particles through characteristic radio signals (see discussion in Meyer-Vernet et al, 2009) measured by the SWAVES instrument (Bougeret et al, 2008) and through trails of debris (St Cyr et al, 2009) imaged in the suite of cameras making up the SECCHI instrument package (Howard et al, 2008).

The SWAVES instrument comprises three orthogonal dipole antennas that are used to measure radio waves generated by solar flares and shocks. These antennas can detect dust grains impacting the spacecraft or antennas via the voltage pulses produced by plasma clouds created by impact ionisation.

The SECCHI package on each spacecraft includes a Heliospheric Imager (Eyles et al, 2009). This is an instrument containing two wide-field white-light cameras mounted on the side of the spacecraft to image any transients in the solar wind travelling along the Sun-Earth line. The sunward of the two HI cameras, HI1, has a $20^o$ field of view with a bore-sight angle $14^o$ from Sun centre in the ecliptic plane. The outer camera, HI2 has a $70^o$ field of view centred at $53.5^o$ from Sun centre. Between them, these two cameras can image the ecliptic from $4^o$ to $88^o$ from Sun-centre. Solar wind transients are imaged through Thomson scattering of sunlight by electrons in the solar wind plasma. Each image that is down-linked to Earth is the sum of many shorter individual exposures that have each been scrubbed for cosmic ray tracks before summation. The dominant signal in an unprocessed HI image results from sunlight scattered by interplanetary dust (which makes up the F-corona). In order to image the much fainter solar wind transients, enough exposures are summed to ensure that the uncertainty in the (relatively constant) F-corona signal is lower than the intensity of any solar wind transient. In this way, the F-corona can be subtracted from HI images either through differencing consecutive images or by subtracting an average F-corona (creating 'background subtracted' images). In synoptic operation, each summed HI1 image represents an exposure of 30 minutes summed over a 40 minute interval while each HI2 image is summed over 2 hours.

With such lengthy summed exposures, any debris in the HI images resulting from the impact of interplanetary dust on the spacecraft appears as long, bright trails usually emanating from a single direction (indicating the point of impact on the spacecraft of that primary dust particle). Since the HI instrument on each spacecraft is mounted on its Earthward-side, each faces in a different direction with respect to the orbital motion of the spacecraft (figure 1). The amount of debris produced from a dust impact is proportional to the kinetic energy of the particle relative to the spacecraft which in turn is presumably proportional to the difference in their velocity squared. Therefore smaller particles can produce debris trails if they have high speeds relative to the spacecraft. Previous studies (St Cyr et al, 2009) have shown that HI on STEREO-A (HI-A) sees many more secondary debris trails than HI-B, indicating that most dust particles impact the spacecraft from the direction towards which the spacecraft is moving. A particle approaching the STEREO-A spacecraft from ahead is more likely to create secondary particles that are ejected into the HI field of view since their momentum will carry them into the lee of the spacecraft where the HI instrument is situated. Conversely, the HI on STEREO-B would most likely see debris resulting from particles impacting the spacecraft from the direction opposite to that in which the spacecraft is orbiting. While debris trails are seen in both the HI1 and HI2 cameras (figure 2), they are more prominent in the HI1 cameras and so the analysis in this paper will concentrate on trails seen in the HI1 cameras only. Similarly, while there is evidence of pointing offsets in the HI2 cameras they are not as clear as in the HI-1 cameras and so only the HI-1 offsets are considered in this paper.

That HI-B experiences more direct particle impacts than HI-A is supported by the difference in pointing stability of the two HI instruments. HI on the STEREO-B spacecraft undergoes frequent transient offsets in the pointing of the cameras, frequently exceeding one pixel (70 arc seconds in HI1) in magnitude. These offsets are calculated from the observations of sudden jumps in the position of background stars used to determine camera pointing. It is thought that these offsets result from the impact of dust onto the HI instrument itself since there are no equivalent offsets seen at the same times in the other SECCHI cameras. HI-B detects more of these impacts since it is facing the direction in which the spacecraft is moving. HI-A, being in the lee of the spacecraft (figure 1), is subjected to far fewer direct impacts. The magnitude of each offset is a convolution of the particle's momentum relative to the spacecraft, the flexibility of the instrument superstructure and the position at which the particle hits the instrument. The most massive particles impacting the edge of the instrument at a point furthest from the mounting points would be expected to generate the greatest moment. Without any information on the exact point of impact for each dust particle, it is difficult to use the magnitude of offset to infer anything about the sizes of impacting particles.

The proximity and speed of the secondary particles imaged by the HI cameras means that most particles rapidly cross the field of view, contaminating a given pixel for no more than a single 40 second exposure. As a result, they can be mistakenly identified as cosmic ray trails by the on-board software and substituted for the uncontaminated value from a previous image. The number of pixels replaced in each exposure by the cosmic ray scrubbing routine is returned within the telemetry and this enables the time of each impact to be accurately determined to within 40 seconds. St Cyr et al (2009) showed that, for HI images containing more than ten debris trails there was a high correlation between the occurrence of dust impacts as measured by the SWAVES instrument and abnormally high values of scrubbed pixels in single HI exposures. They concluded that dust impact was the most likely cause for this subset of HI images containing 'storms' of particle trails.

Each of the previous studies using STEREO data to investigate interplanetary dust have highlighted the distribution of dust particles in different mass regimes. St Cyr et al (2009) detected particles larger than 10 μm from the correlation between debris trails and very large electric pulses detected by the SWAVES Time Domain Sampler. Meyer-Vernet et al (2009) discovered fast (~300 kms$^{-1}$) interplanetary nano particles (0.01 μm) producing electric pulses observed on the low-frequency receiver while Zaslavsky et al (submitted 2011) detected interplanetary beta-particles and interstellar dust, again from the pulses detected by the SWAVES Time Domain Sampler. In the current study, debris trails and offsets in STEREO HI images were used to investigate the distribution of interplanetary dust as an average detected flux per solar longitude at distances of between 0.96 and 1.04 AU from the Sun. By doing so we attempt to identify their source and estimate the mass distribution of the particles.

**2. METHOD**

In this paper, dust impacting the STEREO spacecraft is detected by observing the light reflected from secondary particles (such as flakes of the multi-layer insulation material) as they drift through the HI field of view and, in addition, through offsets to the camera pointing due to particles impacting the instruments directly. Images containing secondary particle trails typically occur several times each day, and their appearance can be both dramatic, containing many bright trails, or more subtle, containing fewer, broader (presumably close, out-of-focus) trails. A selection of HI1 images showing debris trails is shown in figure 2. Most debris enters the field of view from the sunward side of the spacecraft (the right-hand side for HI-A and the left-hand side for HI-B). The insulation material on this sunward face of the spacecraft is a different type from the rest of the spacecraft and appears to be particularly vulnerable to impacts. While most trails are seen in HI-A, one example from HI-B is shown (bottom left). This too shows trails entering from the sunward side of the image.

Identifying such trails in the data is time consuming. With the four HI cameras producing a total of 100 images each day, the number of images to examine after nearly four years of science operations is in the region of 150,000. This task is ideally suited to enthusiastic amateur scientists who are willing to spend time scrutinising images for such trails.

As part of a public engagement project, the STEREO HI team collaborated with the Royal Observatory, Greenwich and the Zooniverse team (Smith et al., 2011) to create Solar Stormwatch (www.solarstormwatch.com). This web-based interface encourages interested members of the public to study images from the STEREO HI cameras in order to detect the occurrence of coronal mass ejections in both the science and near real-time data streams. Solar Stormwatch is proving very popular, attracting over 14,000 active participants. This large group of data analysts are first given some training in each task they perform. Once they have shown that they understand what is being asked of them (by passing a short test) they are given access to background-subtracted images from the inner HI-1 cameras (in the form of movies) served through the web interface. No date or time is shown in each movie clip in order to minimise any preconceptions each member may have about the data. As part of this project, members were asked to identify the frames in which debris trails were seen in the movies. From this information we were then able to generate a database containing all the times of images contaminated by debris as selected by the members. It is possible to estimate the orbital location at which the spacecraft observed these trails to an accuracy of ± 40 minutes. As each STEREO spacecraft orbits the Sun, it becomes possible to map out the spatial distribution of these debris trail observations throughout this orbit.

The advantage of so many people studying the data simultaneously is that multiple identifications of each contaminated image provides greater confidence that the identified time is associated with a genuine event and limits the influence of mistakes or malicious identifications.

While some Stormwatch members have shown a particular aptitude for detecting the fainter trails, such an identification technique will invariably be biased towards the brighter, more spectacular events and be dominated by identifications of trail 'storms'. In order to check the efficiency of this technique, the crowd-sourced data were compared with the trails identified by a single expert (O. C. St Cyr). These expert identifications were carried out using differenced images which, while having a more abstract appearance, can reveal fainter tracks.

**3. RESULTS**

Three sets of results are presented here for comparison: (1) the combined observations of the Solar Stormwatch volunteers looking for debris trails in background-subtracted images, (2) the observations of a single expert looking for debris trails in differenced images and (3) the occurrence of unexpected changes in camera pointing due to primary dust impacts on the HI instruments themselves.

**3.1 Stormwatch results**

Background-subtracted STEREO HI movies since near the start of the mission (Jan 2007) were made available to the Solar Stormwatch community in February 2010. Once the data set had been scrutinised for around eight months, the times of the frames identified as containing debris trails were extracted from the data set. In setting a threshold for the number of identifications required before any identification was considered as reliable, a balance had to be found between setting this threshold as high as possible to ensure reliability and yet not setting it so high that it severely restricted the number of events that were identified.

Several values of this threshold were considered and, while the chosen value altered the overall number of positive identifications, it was found that the exact value did not significantly influence the shape of the distribution of debris images throughout the spacecraft orbit. Thresholds below 5 risked increasing the number of inaccurate identifications while those above 20 served only to restrict the number of positive identifications. Thus, for the purposes of the current paper, a threshold of 10 was used.

For each frame in which trails were observed, the time and date at which that image was taken were used to calculate the position of the spacecraft with respect to the celestial sphere. The number of observations within ten degree bins was then plotted against solar longitude ($0°$ solar longitude being equivalent to $180°$ right ascension).

In order to account for missing data, orbital speed and possible changes to the image cadence throughout the mission, the total number of images within each bin was also calculated. The number of debris trails per bin is presented as a fraction of the total number of images per bin. In this way, the data are not influenced by (small) data gaps and the data sequence does not have to be limited to whole orbits.

While the secondary debris trails resulting from dust impacts were observed from the very earliest stages of the mission, the data presented here are restricted to the time interval between 1$^{st}$ April 2007 and 6$^{th}$ February 2010. This represents the period between the start of science operations and the final date in the initial Stormwatch data set.

The resulting distributions of dust impacts inferred from Solar Stormwatch identifications of debris trails in HI1A and HI1B images over more than three complete orbits of the STEREO spacecraft are presented in figures 3a and 4a respectively. The distribution seen by each spacecraft is quite distinct. For STEREO-Ahead, on which the HI1A instrument is in the lee of the spacecraft, many more images are identified as containing secondary particle trails. The most notable features within the HI-A distribution are the two broad (± 50$^o$) peaks centred on 0$^o$ and 180$^o$ longitude, and the distinct minimum between 240$^o$ and 280$^o$ longitude. In contrast, far fewer HI1B images containing secondary particle trails were identified. The distribution of dust impacts observed throughout the orbit of STEREO-Behind differs markedly from that seen on the Ahead spacecraft, with a single peak in impacts at a solar longitude of around 240$^o$. A reduced number of trails is to be expected for the Behind spacecraft since the HI instrument faces the direction of orbital motion and dust swept up by the spacecraft would generate secondary particles that would mostly propagate away from the HI field of view. Poisson statistics have been used to estimate the uncertainties in each longitude bin and errorbars representing one standard deviation are displayed on each bin. In order to examine the significance of the peaks and troughs within each distribution, the mean value for all bins is also plotted as a horizontal dashed line. For the distribution of trails observed with HI1A (figure 3a), only the maximum at 0$^o$ and the minimum between 240$^o$ and 280$^o$ fall beyond one standard deviation from the mean while the same is true for the central peak in the distribution seen by HI1B (figure 4a). Even with such low numbers of events, there are hints that the distributions observed by both spacecraft are not entirely random in nature. In order to test for this, a runs test was applied to each distribution. This compared the observed distribution against the null hypothesis that the values in the distribution are in random order. The runs test indicated that the Stormwatch distributions in both HI1A and HI1B (figures 3a and 4a) are not statistically distinct from a random distribution due to the relatively low number of events seen in each longitude bin. These distributions, while noisier, are however consistent with the distributions obtained by a single expert, for which there are approximately twice the number of observations.

**3.2 Single expert results**

Secondary particle trails observed in differenced images were also collated by a single expert observer. The data set that was used in this study spanned the interval from 1$^{st}$ April 2007 to 31$^{st}$ August 2010. The orbital positions of the two spacecraft were also calculated for the times of these observations and, again, these were summed into 10$^o$ bins and presented as a fraction of the total number of image per bin. This enables a direct comparison with the results from the Stormwatch observers. The results for H1-A are presented in figure 3b while those for HI-B are presented in figure 4b. The distribution presented in black corresponds to all images containing trails while that in white represents the subset of images containing 'storms' of trails (more than ten per image). The most striking feature when comparing the HI-A results in figures 3a and 3b is that the single expert has made far more (over double) positive identifications of dust trails in the images. There may be several reasons for this. The most likely is due to these trails being identified from differenced images which, while being more complex to interpret, are better able to show up fainter tracks. The number of trails indentified by the Stormwatch community is also limited by the threshold of independent identifications required. Lowering this threshold increases the number of trails indentified but the total number is still below that observed by the single expert. It is reasonable to conclude that the Stormwatch identifications are a subset of the total population, containing images with the brightest trails or those with multiple trails.

Although the overall number of images in which trails were identified differs between the amateur and expert surveys of HI1 images from the A spacecraft, the spatial distributions are very similar. Figure 3b again contains two broad (±50°) peaks centred on 0° and 180° with a (less) distinct minimum between 240° and 270°. There is also a broad shallow minimum between 30° and 80° which also appears (although slightly less prominently) in the Stormwatch distribution plotted in figure 3a. The images with trails identified by a single expert in the HI-B data (figure 4b) compared with those identified by the Stormwatch community (figure 4a) reveals a distribution that also has a prominent asymmetric peak centred around 240°, and also a secondary broad (±50°) peak centred on 0°. There are also slight indications of minima around 120° and 300°.

The increased number of events identified by a single expert for HI1A (figure 3b) results in a less noisy distribution compared with those identified by Solar Stormwatch volunteers (figure 3a). For the distribution in figure 3b (black), both minima only briefly exceed one standard deviation from the mean while the central peak is within one standard deviation throughout. While the variation of counts with longitude does not deviate significantly from the mean, a runs test confirms that there is less than a 5% chance that such a distribution is random. When the same tests are applied to the distribution of events identified by a single expert in the in HI1-B data (figure 4b), the maximum at 240° and the minimum at 120° both lie beyond one standard deviation from the mean while the runs test suggests that there is around a 10% chance such a distribution is purely random.

It is interesting to note that, for the expert identifications from both HI-A and HI-B (figures 3b and 4b), the distribution of all images containing trails (black) differs markedly from the distribution of images containing 'storms' of trails (white). If the occurrence of images containing 'storms' is subtracted from the overall distributions for each spacecraft, the resulting distributions of 'non-storm' events (i.e. those corresponding to images with less than ten debris trails) around the STEREO orbits are revealed (figure 5). It can be seen that the distribution is quite different for each spacecraft. For HI-A, there is a broad peak at solar longitudes around 180° and a secondary peak around 0°. For HI-B there is a narrow peak at 250° and minima around 120° and 220° in solar longitude. While an unknown but potentially sizeable proportion of these events may be due to the continuous degradation of the spacecraft insulation material and not due to particle impacts, such degradation would not be expected to vary significantly throughout the orbit or be so different for the two spacecraft.

The distribution of HI-A 'storms' (figure 3b: white) was correlated with the 'non-storm' distributions from both HI-A (figure 5a) and HI-B (figure 5b). It was found that the correlation coefficients were -0.62 and 0.37 respectively, indicating that the distribution of 'storms' seen in HI-A more closely matched the 'non-storms' in HI-B. Similarly the distribution of HI-B 'storms' (figure 4b: white) was correlated with the 'non-storm' distributions from both HI-A (figure 5a) and HI-B (figure 5b). The correlation coefficients were found to be 0.45 and -0.41 respectively. The distribution of 'storms' seen in HI-B more closely matches the 'non-storm' distribution in HI-A.

Further, the 'non-storm' distribution for HI-A (figure 5a) is most similar to the off-point distribution in HI-A (figure 4c) while the 'non-storm' distribution for HI-B (figure 5b) is most similar to the distribution of off-points in HI-B (figure 3c). These off-point results will be discussed in the next section.

These results are consistent with the variation around the STEREO orbit of 'non-storm' trails being due to particles impacting each spacecraft having approached from the same side as the HI instrument. This would lead to secondary particle trails that mostly fall outside the HI field of view (although a small number may be observed) as the momentum of the primary particle is transferred to the secondary debris. The same population of primary particles would generate off-points in the HI images. On the other spacecraft, the same population of primary particles would generate secondary trails that would mostly tend to move through the HI field of view as momentum was again conserved.

For HI-A 'storm' trails, HI-B offsets and HI-B 'non-storm' trails, this would require particles to approach from the direction towards which the spacecraft is moving (the apex direction of the spacecraft) while for HI-B 'storm' trails, HI-A offsets and HI-A 'non-storm' trails, this would require particles to approach from opposite (anti-apex) direction of the spacecraft.

**3.3 Dust impact results**

In a similar way, it is possible to map the distribution around the spacecraft orbits of transient pointing offsets that sporadically occur between adjacent HI images, thought to be due to dust particles striking the instrument. The pointing direction of the HI instruments is determined to a high degree of accuracy by fitting the observed star-field to a star catalogue (Brown et al, 2009). This is done independently for each image and so it is possible to determine the difference in pointing between consecutive images. Images from the inner (HI1) cameras were used for this study since off-points are more noticeable in their smaller field of view compared with the outer (HI2) cameras. Setting a threshold above which such off-points are considered meaningful is difficult when the magnitude of an off-point is a convolution of particle momentum and the position at which the particle strikes the instrument. There are many more small pointing offsets (less than a pixel) than there are larger ones. Figures 3c and 4c present the distribution of these offsets in HI-B and HI-A, respectively, with values exceeding thresholds of ; 0.5, 1, 2 and 4 pixels. To enable direct comparison with the other panels in figures 3 and 4 these have been summed into bins with a width of $10^o$ in longitude and are again presented as a fraction of the total number of images per bin. Since impacts on HI-B are expected to result from dust particles that have a component of their velocity in the opposite direction to the spacecraft, the distribution of these impacts should match the distribution of trails seen in HI-A since they are likely to be caused by particles with similar orbital characteristics. Conversely, the distribution of impacts seen on HI-A should match that of 'storm' trails seen in HI-B.

This is why the distribution of pointing offsets for HI-B is plotted with the distributions of HI-A dust trails in figure 3 and the offset distribution for HI-A is plotted below HI-B dust trails in figure 4.

For HI-B (figure 3c) the distribution of off-points appears to be similar for all chosen values of the off-point threshold, though the number of off-points varies markedly with the chosen threshold value. For pointing offsets exceeding a threshold of 0.25 pixels (black shading) the distribution shows a prominent maximum at around $0°$ with the hint of another peak around $180°$. The most noticeable difference between this and higher thresholds however is the appearance of two additional but narrow peaks around $80°$ and $260°$, making the identification of minima more difficult. Since the number of instrument off-points in HI-B is much greater than the observed number of particle trails, the uncertainties in each longitude bin are much smaller for the off-points compared with the particle trails. For HI-B all features lie well outside one standard deviation from the mean. For HI-A, where the number of off-points is lower, the only feature to significantly deviate from the mean is the peak at around $240°$ longitude (figure 4c). The number of observed impacts depends on the off-point threshold applied but it is possible to estimate the approximate mass of these particles by considering estimates of the dust distribution at 1 AU (Grün et al, 1985). For HI-B, approximately 10% of images experience an off-point larger than 0.25 pixels at a solar longitude of $180°$. At this point in its orbit, STEREO-B travels through $10°$ of solar longitude in around 11 days. Since the HI instrument has a surface area of approximately 0.28 $m^2$, and records 36 images per day, this represents a total flux of approximately $2.6 \times 10^{-4}$ $m^{-2}s^{-1}$ which equates to particles with masses in excess of $10^{-17}$ kg and sizes exceeding 0.1 µm (Grün et al, 1985).

For HI-A (figure 4c) there are far fewer off-points identified than for HI-B. Since the HI-A instrument is in the lee of the spacecraft, the off-points it experiences cannot be due to impacts with particles travelling from the direction towards which the spacecraft is moving since the instrument is shielded from these by the spacecraft itself. Since the distribution of these off-points seen along the orbit of STEREO-A is consistent with the distribution of 'storm' trails observed in the HI-B images, and this distribution has a distinct peak at a longitude of $250°$, it seems reasonable to conclude that this peak is also caused by particles with a component of their orbit in the same direction as the spacecraft motion.

**4. COMPARISON WITH METEOR OBSERVATIONS AT EARTH**

Since the STEREO spacecraft are in heliocentric orbits at similar distances from the Sun as the Earth, it may be expected that the distribution of dust detected by the spacecraft along their orbits would be similar to the influx of meteors arriving at Earth. This comparison assumes that the particles impacting the STEREO spacecraft and those generating observable meteor trails in Earth's atmosphere are from the same population and, also, that this population contains particles with sufficient size and speed (relative to the Earth and the spacecraft) to generate both debris trails and off-points in the HI images and observable meteor trails in the Earth's atmosphere. If the particle mass estimated from the instrument off-points is correct ($> 10^{-17}$ kg), the smaller particles from this distribution are not of sufficient size to generate meteor trails at Earth. For example, the limiting mass for the ground-based Canadian Meteor Orbital Radar (CMOR) is $4.10 \times 10^{-7}$ kg. It has not been possible to estimate the mass of the particles from the occurrence of secondary particle trails seen in STEREO/HI data however and so a comparison of the occurrence of HI particle trails with the meteoroid population is potentially useful.

The population of meteoroids known to produce meteor trails at Earth can be divided into two broad categories; those that orbit in narrow streams often associated with a parent comet or asteroid and those that do not belong to any of these streams, which are known as sporadic meteoroids. Much attention is given to the narrow streams since these can generate spectacular, if short-lived, meteor showers at Earth but the majority of meteoroids that strike the Earth actually come from the sporadic meteor population. The latter are thought to be comprised of both particles generated by inter-asteroidal collision and cometary particles deflected by gravitational and radiative forces from their original narrow orbits into a diffuse population (Wiegert et al, 2009). The sporadic background is not without structure, however, with several broad populations that are observed to come from six well-defined directions relative to the Sun (e.g. Campbell-Brown and Jones, 2006). The north and south apex sources are centred on the apex of the Earth's way (towards which the Earth is moving) around $15°$ above and below the ecliptic respectively (Sekanina, 1976). The helion and antihelion sources (Hawkins, 1956; Weiss and Smith, 1960) are located in the ecliptic between $60°$ and $70°$ from the Earth's apex direction while the north and south toroidal sources (Elford and Hawkins, 1964; Jones and Brown, 1993) are located approximately $60°$ north and south of the ecliptic.

Daily rates for these meteors arriving at a particular location on Earth are modulated by the changing position of the observing station relative to these sources. Diurnal meteor rates are greatest around local dawn while the annual rates are greater in the second half of the year in the northern hemisphere. For a given observing station, these factors can be corrected for, allowing cross-comparison of meteor rates measured globally.

For this current study, meteor observations were taken from the Canadian Meteor Orbital Radar (CMOR, Jones et al, 2005) located in Tavistock, Ontario, Canada. Since the orbital characteristics of meteors associated with known streams are well defined, these can be identified and separated from those associated with the sporadic meteor population. After correcting for observing biases, the meteor fluxes are well defined in a relative sense but the individual fluxes have been scaled to an average of unity. Since we are concerned with comparing the variation in sporadic meteor fluxes around Earth's orbit with the effects of impacts seen on the STEREO spacecraft, this is more than sufficient for our purposes.

Four of the six sporadic meteor streams observed with the CMOR radar between April $1^{st}$ 2007 and August $31^{st}$ December 2008 (corrected for observing biases) are plotted in figure 6. In order to aid comparison with the STEREO data, the meteor data were again binned into $10°$ bins and their mean value plotted against solar longitude. The Poisson errors in the resulting distributions are not shown but they will be of the order of a few percent since the CMOR radar detects thousands of meteoroids per day. Of the six sporadic sources, the south toroidal and south apex sources are the least well sampled due to the northern latitude of the observing station and are not presented here. From a simple comparison with figures 3 and 4, it is apparent that while the north toroidal and north apex sources show the closest match to the broad distribution of the total debris trails seen in the HI images (figures 3a, 3b and 4a, 4b, black), the distribution of 'storm' events shares many of the features of the Helion source.

There are two prominent spikes in the north toroidal and north apex distributions at around $300°$ and $120°$ longitude in the CMOR radar data. The spike at $300°$ could be contamination from the

Quadrantid meteor shower. While meteors with radiants far from the source being considered have been removed from these plots of sporadic sources, there are so many Quadrantid meteors that some cross-contamination can occur. Similarly the spike at 120$^o$ corresponds to the time of the psi Cassiopeids and alpha Lacertids meteor showers from which some cross–contamination could have resulted. There are two similar peaks seen in the HI-A debris trails shown in figure 3a (corresponding to images with the most prominent trails) but the longitudes of these peaks do not coincide exactly with those of the same meteor streams mentioned above. Since the STEREO-A spacecraft orbits at a distance of 0.96 AU it could just be that it does not encounter the same part of these relatively narrow meteoroid streams or that the streams intersect the orbits of Earth and the STEREO spacecraft at slightly different locations. These two meteor streams could also be responsible for the additional similarly located peaks in off-points of the HI-B instrument shown in figure 3c although the latter are much broader than those seen in the meteor data. If they do prove to be from the same source, this could provide information about the size distribution of particles within these meteor showers.

## 5. ESTIMATING THE MASS DISTRIBUTION OF DUST PARTICLES FROM HI OFF-POINTS

Sporadic sources near the apex direction of the Earth's way will have large speeds relative to the Earth (e.g. Galligan and Baggaley, 2002) and the two STEREO spacecraft and so it is perhaps unsurprising that the distribution of impacts seen on the STEREO spacecraft should match these sources the most closely.

There is some evidence for different sources dominating in different mass regimes. Very high-power radar observations of meteoric material at Earth, which correspond to very small (close to 10 μm) particles, tend to be dominated by the apex sources, while observations in the 100 μm - 1 mm size range are dominated by the helion and antihelion sources. There seem to be dynamical reasons for this: modelling of the sporadic complex shows that small apex material is more efficiently delivered to the Earth (Wiegert et al, 2009).

While there are some features in both CMOR and STEREO observations that could possibly be due to certain meteor streams, individual meteor streams are not, in general, distinct within in the distributions of impact effects observed on the STEREO spacecraft throughout their orbits. This could be due to a number of reasons. Firstly, these streams are more spatially constrained than the sporadic streams and are much more sensitive to the position of the observer. The orbital radii of the STEREO spacecraft differ from the orbital radius of the Earth by around 0.04 AU (7.5 million kilometres). Secondly, the mass and velocity of particles capable of causing off-points or secondary trails in the spacecraft data may be different from the mass and velocity of particles generating meteor trails in the Earth's atmosphere.

Since both the amount of debris ejected from the STEREO spacecraft and the particle momentum (causing off-points) do increase with speed, small particles can be detected more easily if they have high velocities relative to the spacecraft. As a result, we should detect a larger proportion of small particles from the apex direction than from anti-apex, even if the fluxes are similar. This could introduce a bias when comparing estimates of the mass distributions measured from impacts on each spacecraft.

The mass distribution of material entering the Earth's atmosphere can be estimated from the received power of radio echoes reflected from meteor trails (Hawkins, 1956; Blaauw et al, 2011). The mass distribution within a population of meteors can be investigated using the mass distribution index. The mass distribution is a power law that can be described by the equation;

$$dN_c = cM^{-s} dM \qquad (1)$$

where $dN_c$ is the number of particles between mass $M$ and $M+dM$, c is a normalizing factor and $s$ is the mass index. By plotting the cumulative number of meteoroids versus mass, the differential mass index can be found from the gradient. A differential mass index higher than two indicates that there is more mass in the smaller particles, whereas a differential mass index lower than two indicates that there is more mass in the larger particles. Calculations of the mass index are useful when considering the hazard posed to spacecraft by the individual sporadic sources.

The mass index calculated for each of the sporadic sources measured by the CMOR radar is of the order of 2.1, varying between 2.0 and 2.3 throughout the Earth's orbit (Blaauw et al, 2011). In contrast, most streams producing meteor showers observed at Earth have a mass index less than 2, indicating that they are deficient in small particles. While CMOR sees many shower meteors (e.g. Brown et al, 2010), the Advanced Meteor Orbit Radar (AMOR, which detects particles down to 40 µm) sees just a couple of significant showers (Galligan and Baggaley, 2002), and the High Power Large Aperture Radar (HPLAR) at Jicamarca, which detects the smallest sizes (~$10^{-12}$ kg), is only able to detect meteor showers through interferometric techniques (Chau and Galindo, 2008).

Integrating equation (1) gives

$$N_c \propto M^{-(s-1)} \qquad (2)$$

Where $N_c$ is the cumulative number of meteoroids larger than mass $M$. A logarithmic plot of the cumulative number versus slope therefore has a slope of *1-s*.

While it is difficult to infer any information about particle mass from the secondary trails that are seen in the HI images, the direct impact of dust on the instruments provides a means by which we can estimate the momentum distribution of the particles. For this, we assume that the magnitude of an off-point is proportional to the momentum of the incoming particle and that the impacts on the instrument are randomly distributed. If we further assume that the incoming particles all come from the same population whose velocity is independent of mass, then we can estimate the mass distribution of the impacting particles by studying the distribution of off-points. $N_c$, in this case would represent the cumulative number of off-points greater than a given magnitude (in pixels).

Using the assumptions above, mass index values were calculated for each longitude bin of the off-point distributions presented in figures 3c and 4c by fitting the gradient to plots of $N_c$ versus off-point threshold (assumed to be proportional to mass). The distributions of the mass-index along the STEREO spacecraft orbits for both HI-A and HI-B are presented in figure 7. The difference between the two is marked, with the mass index for the particles impacting on HI-B (facing the ram direction of the spacecraft and therefore exposed to particles from the apex and toroidal sources) being predominantly greater than 2 throughout almost the entire orbit. In contrast, the mass index values calculated for HI-A (in the lee of the spacecraft and exposed to particles approaching from the anti-apex direction) are much lower, ranging between 1 and 1.9. These results suggest that a greater proportion of the mass impacting on HI-B is in the smaller particles, as is the case for meteors observed at Earth (Blaauw et al, 2011), even though the mass range of our detections from off-points is smaller by several orders of magnitude . In contrast, most of the mass from particles impacting on HI-A is in the larger particles. The off-points in HI-A are most likely caused by particles travelling in the anti-apex direction while those in HI-B are caused by particles travelling in the apex direction. Assuming that these two populations have equal and opposite velocity distributions, the relative velocity of the STEREO-A spacecraft with respect to anti-apex particles would be lower than for apex particles approaching STEREO-B. The observed difference in mass indexes measured at the two spacecraft is consistent with the bias introduced by the relative velocity between the spacecraft and the dust particles. This is consistent with the lack of anti-apex meteors observed at Earth, thought to be due to their having small relative velocities insufficient to produce enough ionisation to be detected by radar (Webster et al, 2004).

## 6. CONCLUSIONS

The distribution of dust impacts on the two STEREO spacecraft throughout their orbits have been inferred through observation of secondary particle debris trails and unexpected off-points in the HI cameras. A comparison between observations of the brightest events (observed in background-subtracted images by a community of enthusiastic amateurs) and a survey including fainter trails (seen in differenced images by a single expert observer) shows consistent distributions. While there is no obvious correlation between these distributions and the occurrence of individual meteor streams at Earth, there are some broad features consistent with the helion, toroidal and apex sources of the sporadic meteor population. The toroidal and apex sources both show broad (±50°) peaks around 0° and 180° solar longitude, similar to those seen in HI1-A trails (figures 3a and 3b) while the broad (±50°) peak in the helion source at 270° is similar to that seen in the HI1-B trails (figures 4a and 4b).

The distribution of off-points experienced by each of the HI instruments throughout their orbits most closely resembles the 'storm' debris trails seen in the HI images taken from the other STEREO spacecraft. For off-points of the HI-A instrument, which can only have been caused by particles travelling from the anti-apex direction, the distribution is consistent with the distribution of 'storms' of secondary trails observed in the HI-B camera, providing evidence that these trails also result from impacts with primary particles from an anti-apex source.

Investigating the mass distribution for the off-points observed in HI-B, it is apparent that the mass index of particles from the apex direction is consistently above 2. This indicates that the majority of the mass is within the smaller particles of this population. In contrast, the mass index of particles from the anti-apex direction (causing off-points in HI-A), where it is possible to derive a value, is consistently below 2, indicating that the majority of the mass is to be found in larger particles of this distribution. The anti-apex sources at Earth are more difficult to observe due to their relatively low velocity. Impacts observed by the STEREO spacecraft could provide an insight into the distribution of this source.

**Acknowledgements**

The UK STEREO group is supported by funding from the UK Space Agency. Solar Stormwatch is a collaborative project between the Zooniverse team, the Royal Observatory Greenwich and the Science and Technology Facilities Council. The Zooniverse is supported by The Leverhulme Trust. The STEREO Heliospheric Imagers are supported by the UK Space Agency. STEREO HI Data is made available via the UK Solar System Data Centre (www.ukssdc.ac.uk)

**Figure Captions**

**Figure 1.** A cartoon demonstrating the relative position and motion of the two STEREO spacecraft. The Heliospheric Imager on the Ahead spacecraft (HI-A, represented by the solid black rectangle) is in the lee of the spacecraft and so observes secondary trails resulting from particles travelling in the opposite direction to the spacecraft and impacting in its surface. Conversely, HI-B (also represented as a solid black rectangle) faces the direction of motion of the spacecraft and so particles travelling in the opposite direction to the spacecraft impact on the instrument itself.

**Figure 2.** Examples of secondary trails seen in the HI cameras throughout the period of interest. Most trails are seen to emanate from the sunward-side of the spacecraft (the right-hand side for HI-A and the left-hand side for HI-B).

**Figure 3.** The distribution of dust impacts throughout the STEREO spacecraft orbits identified through secondary particle trails within HI images on STEREO-A (top two panels) and pointing offsets in HI on STEREO-B (lower panel) as a function of solar longitude. The top panel shows events identified from background-subtracted images by members of the Solar Stormwatch project. The total number of frames containing trails is presented as a fraction of the total number of images in each $10^o$ longitude bin. The middle panel shows the distribution calculated from observations by a single expert identifying secondary trails in differenced HI images. All images containing debris trails are shown in black while those images containing ten or more debris trails are shown in white. Once again, these are presented as the fraction of images containing trails within each $10^o$ longitude bin. The bottom panel contains the distribution of pointing off-sets (in units of pixels) seen in HI-B images. Four thresholds were considered; > 0.25 (black), > 1.0 (red), > 2.0 (blue) and > 4.0 (white). For each $10^o$ longitude bin, the number of off-points is expressed as a fraction of the total number of images. Each distribution is plotted with errorbars on each bin and the mean value of all bins in the distribution is shown as a horizontal dashed line.

**Figure 4.** The distribution of dust impacts throughout the STEREO spacecraft orbits identified through secondary particle trails within HI images on STEREO-B (top two panels) and pointing offsets in HI on STEREO-A (lower panel) as a function of solar longitude. The top panel shows events identified from background-subtracted images by members of the Solar Stormwatch project. The total number of frames containing trails is presented as a fraction of the total number of images in each $10^o$ longitude bin. The middle panel shows the distribution calculated from observations by a single expert identifying secondary trails in differenced HI images. All images containing debris trails are shown in black while those images containing ten or more debris trails are shown in white. Once again, these are presented as the fraction of images containing trails within each $10^o$ longitude bin. The bottom panel contains the distribution of pointing off-sets (in units of pixels) seen in HI-A images. Four thresholds were considered; > 0.25 (black), > 1.0 (red), > 2.0 (blue) and > 4.0 (white).

For each 10$^\circ$ longitude bin, the number of off-points is expressed as a fraction of the total number of images. Each distribution is plotted with errorbars on each bin and the mean value of all bins in the distribution is shown as a horizontal dashed line.

**Figure 5.** The distribution of HI images throughout the STEREO spacecraft orbits containing 'non-storm' (less than ten) trails for HI-A (figure 5a) and HI-B (figure 5b) as a function of solar longitude. Each distribution is plotted with errorbars on each bin and the mean value of all bins in the distribution is shown as a horizontal dashed line.

**Figure 6**. The distributions of four sporadic meteor streams throughout Earth's orbit as observed by the Canadian Meteor Orbital Radar. These data were obtained between 2002 and 2009 and as such represent an update to the distributions published previously (Capmbell-Brown and Jones, 2006). Counts have been corrected for observing biases and normalised before being averaged into 10$^\circ$ bins in solar longitude for comparison with the STEREO data. From the top panel downwards these are; north apex, north toroidal, helion and antihelion. The bottom panel contains the distribution of all the sources combined.

**Figure 7.** Estimates of the mass index estimated from the distribution of off-points seen in the HI-A and HI-B cameras as a function of solar longitude. For HI-B, (figure 6a) the mass index is above 2 for most of the orbit, indicating that most of the mass is contained within the smaller particles of the dust population whose impacts are thought to be causing the off-sets. Conversely, the mass index calculated for offsets observed in HI-A images are consistently below 2 where it can be measured throughout the orbit (6b), indicating that most of the mass for the population of particles causing these off-points is contained within the larger particles.

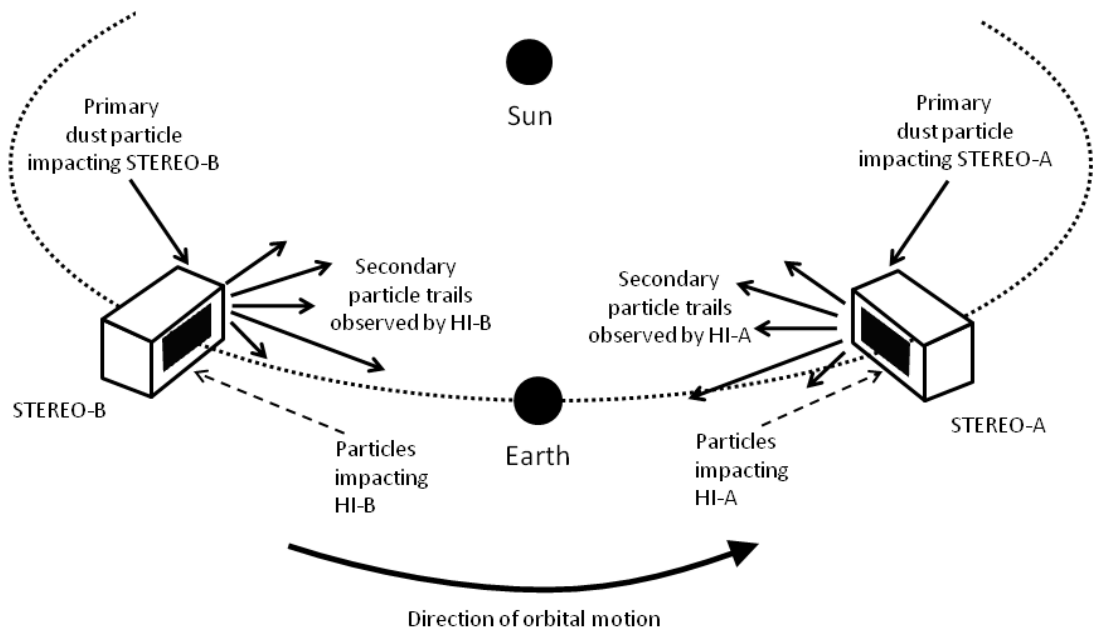

Figure 1.

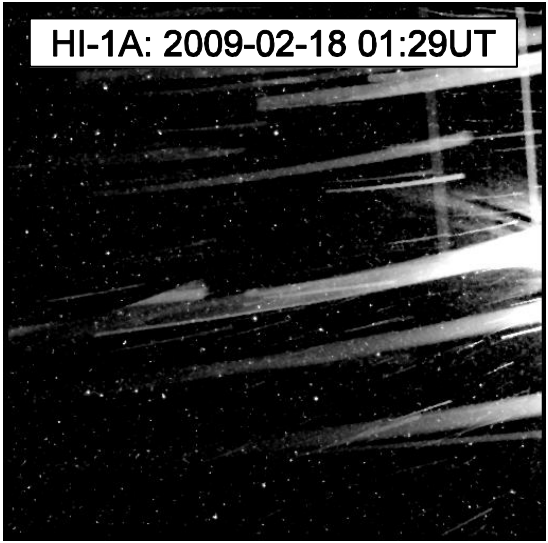
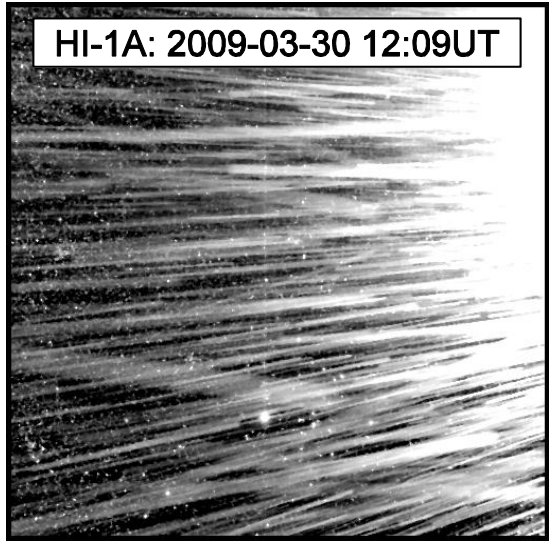
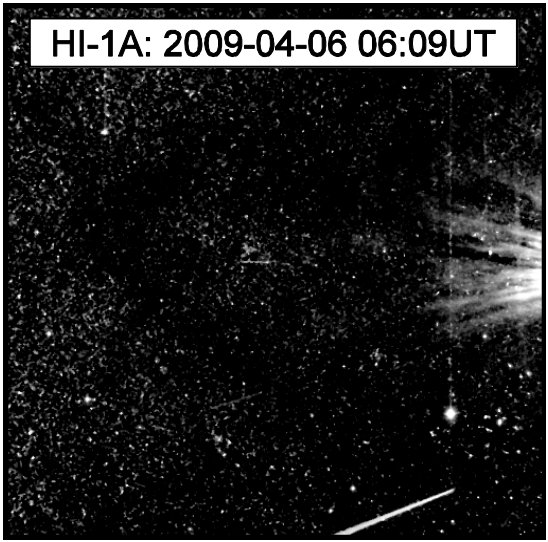
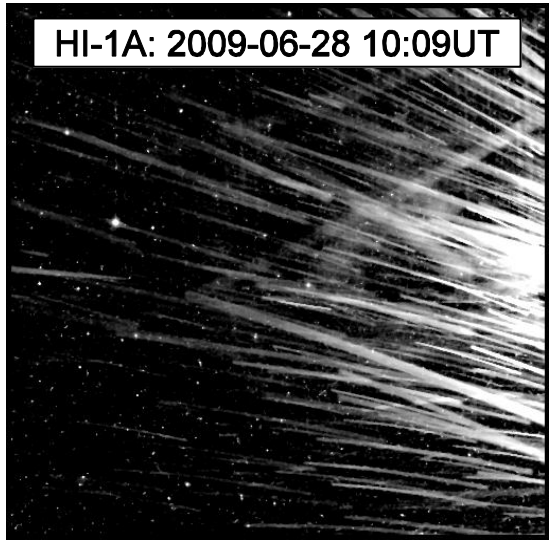
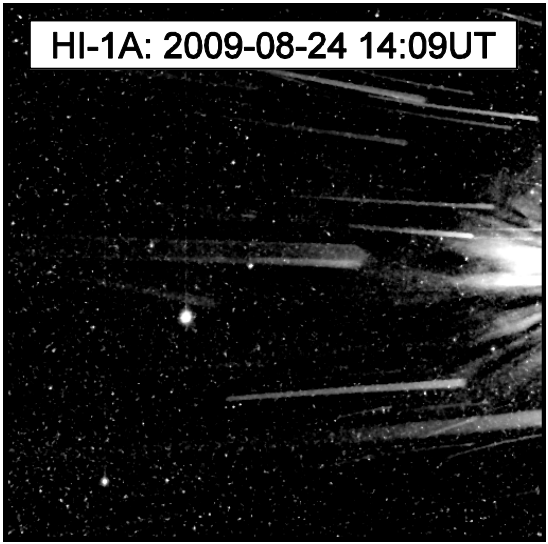
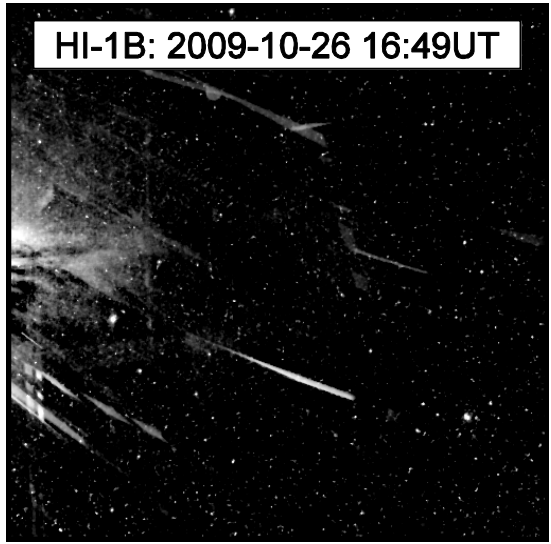

**Figure 2.**

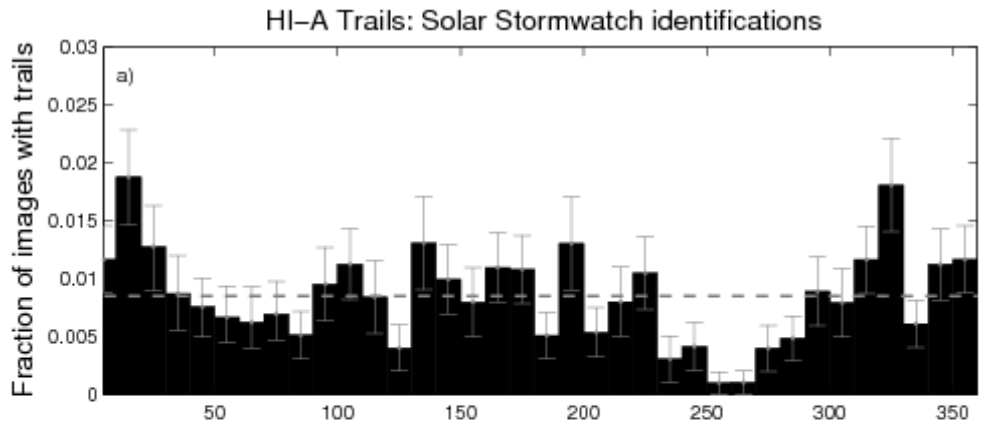
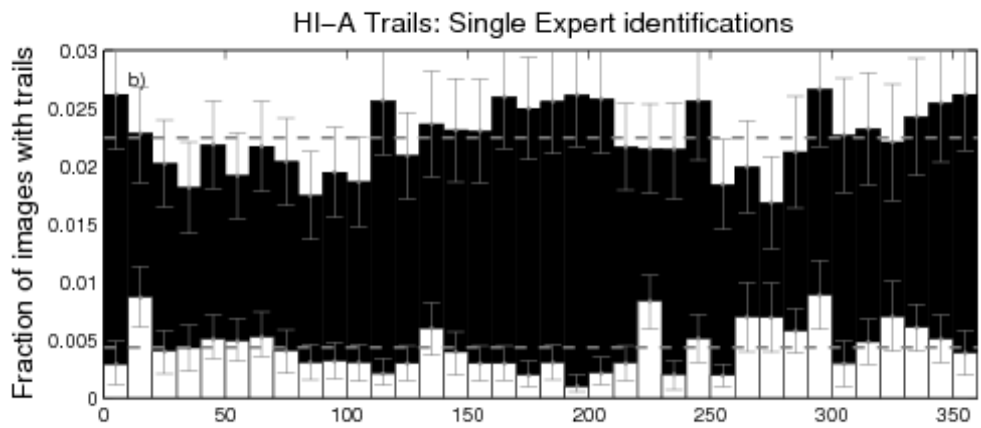
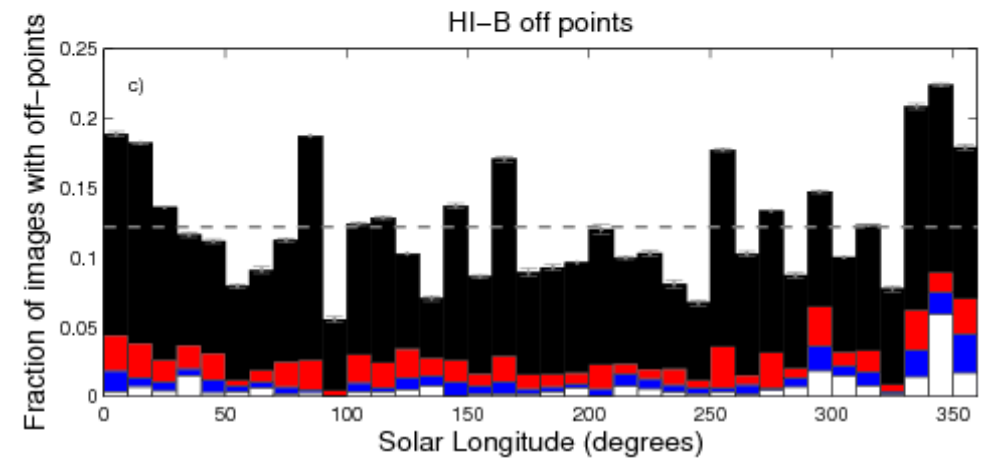

**Figure 3.**

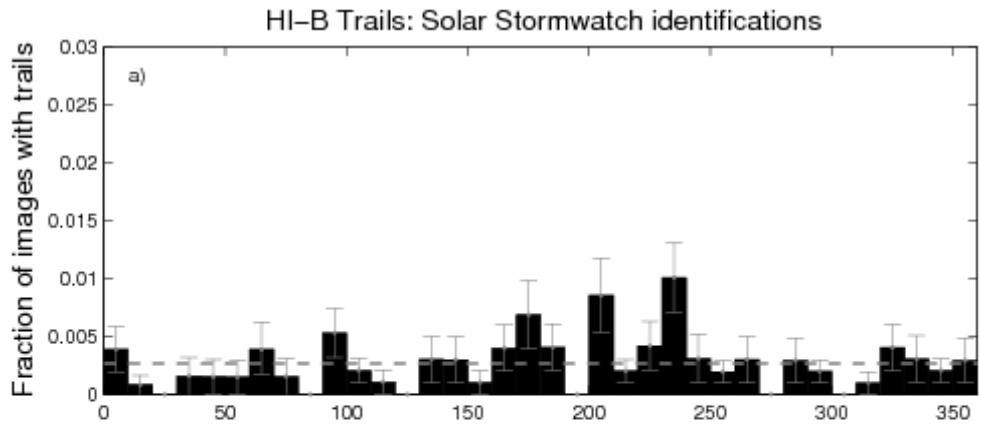

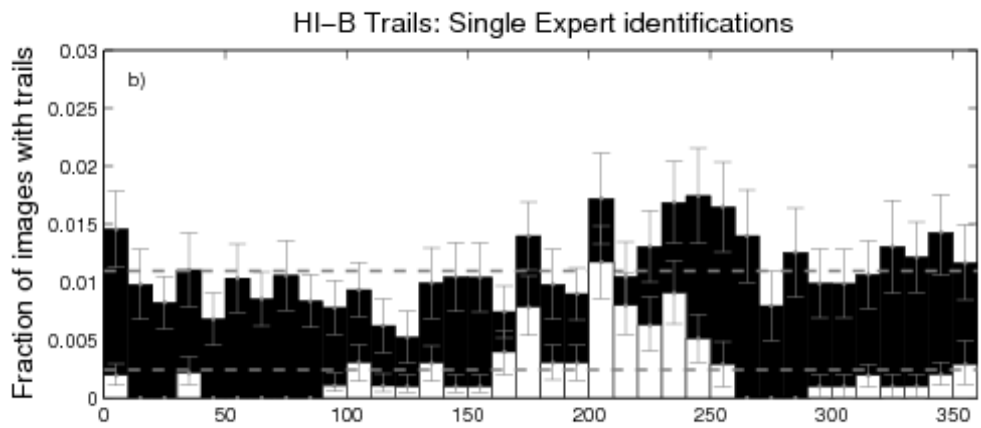

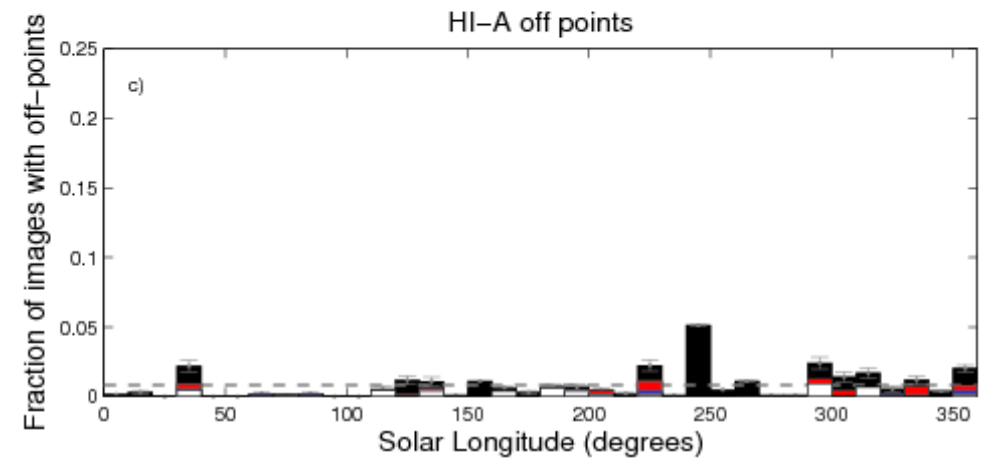

**Figure 4**

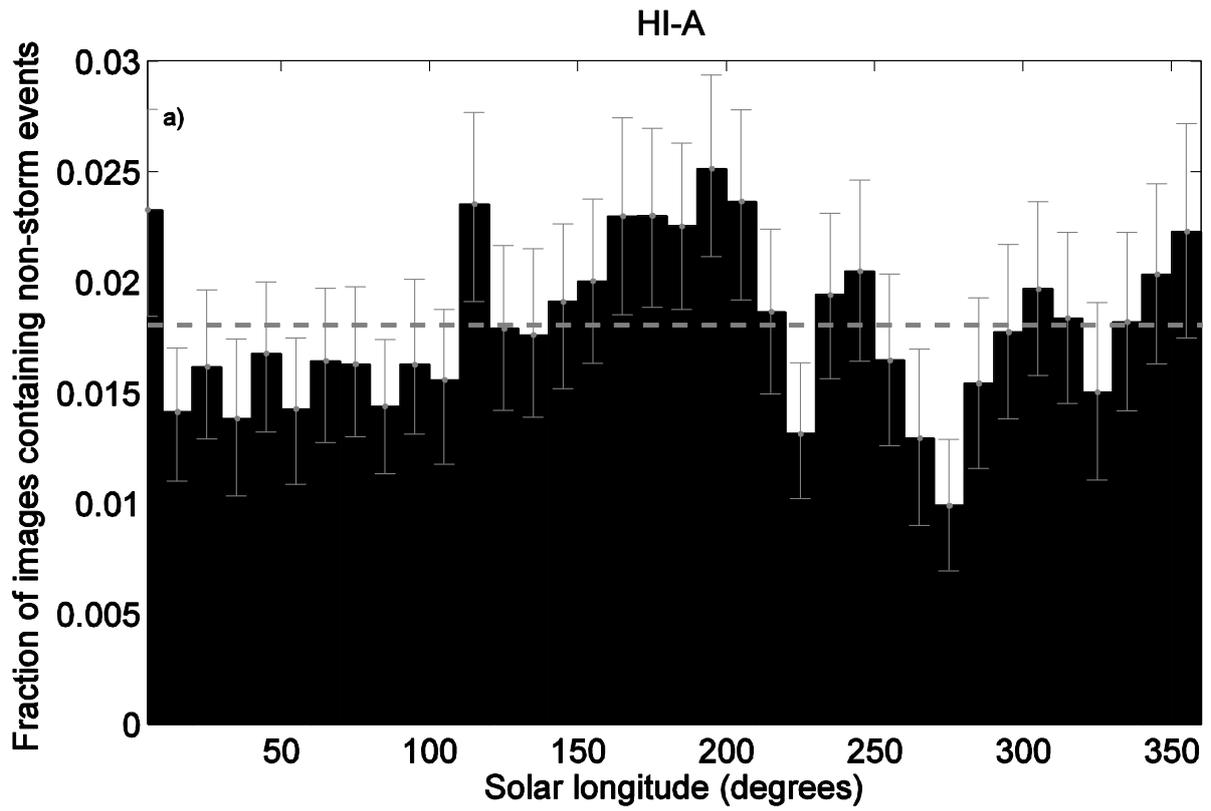
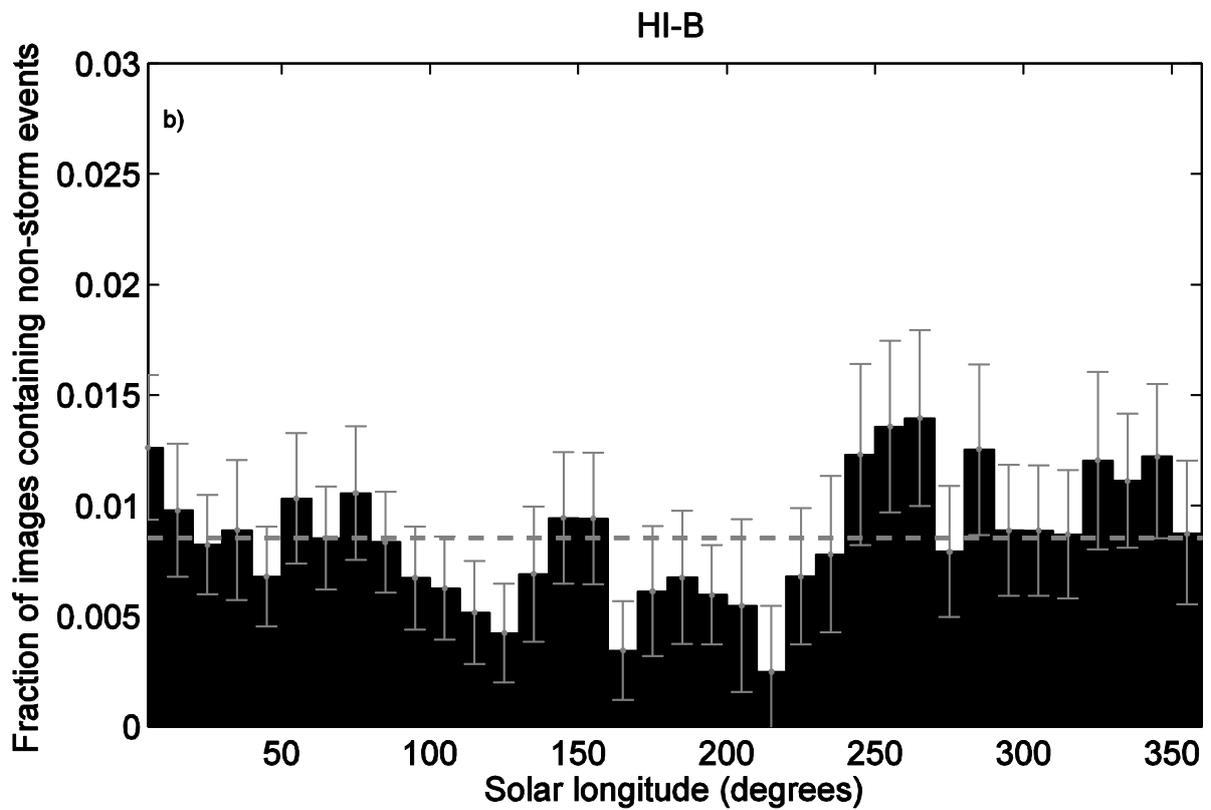

Figure 5

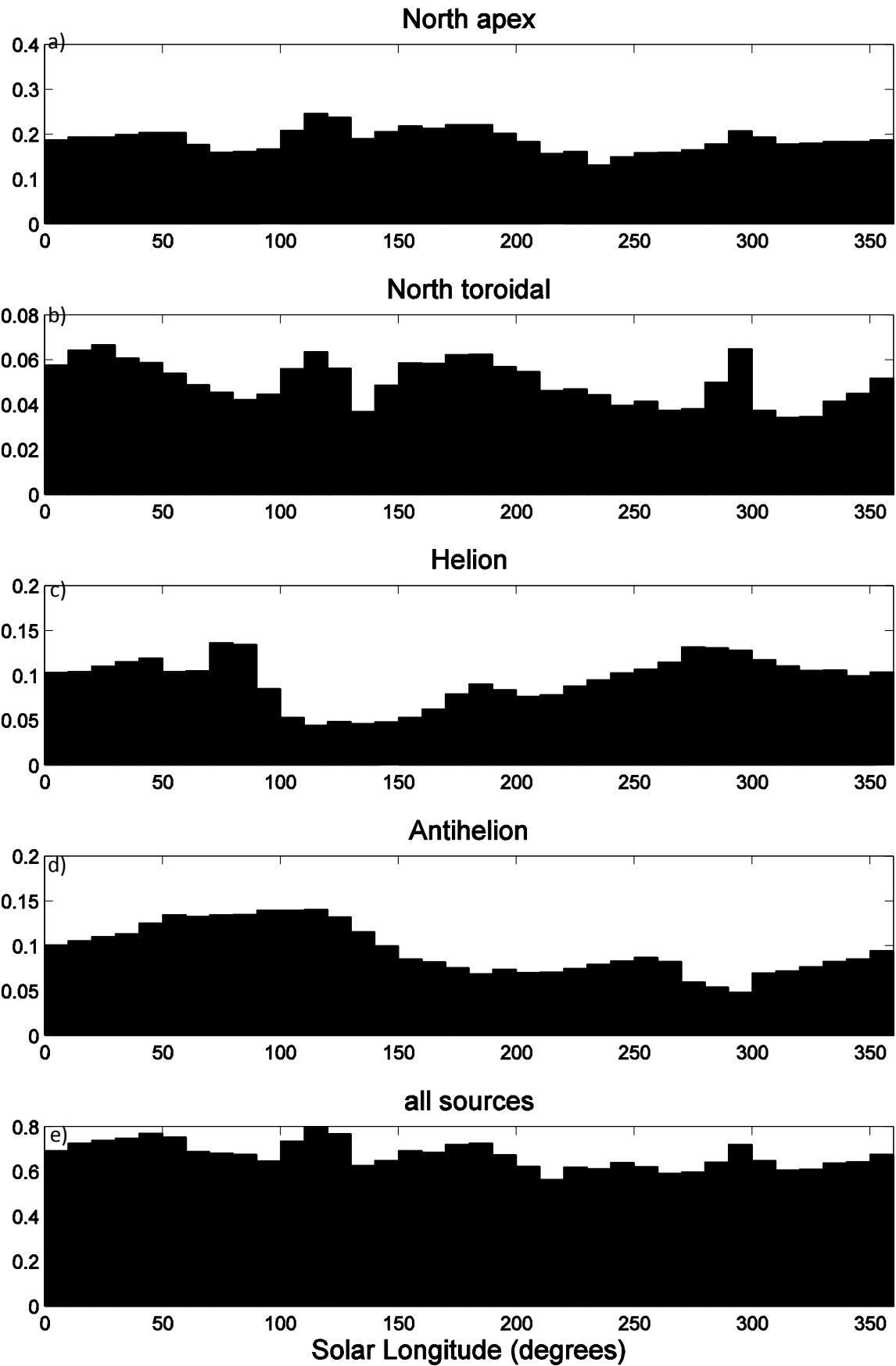

Figure 6

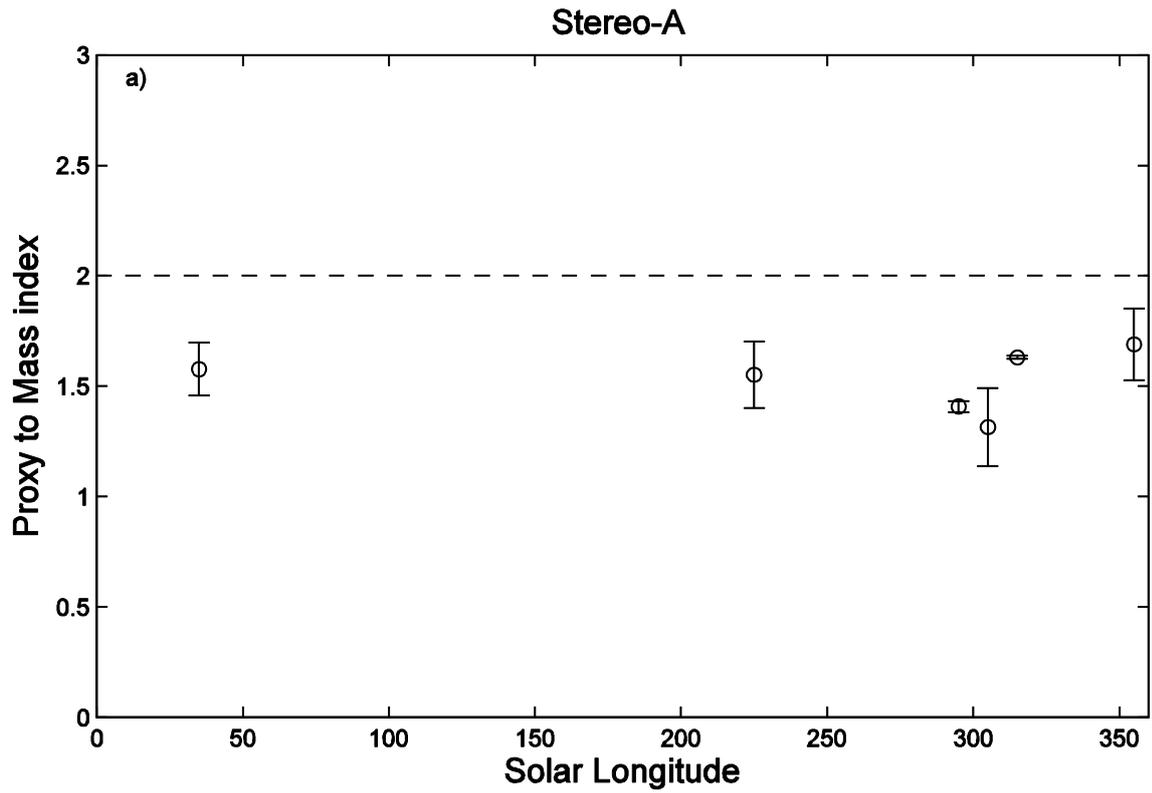

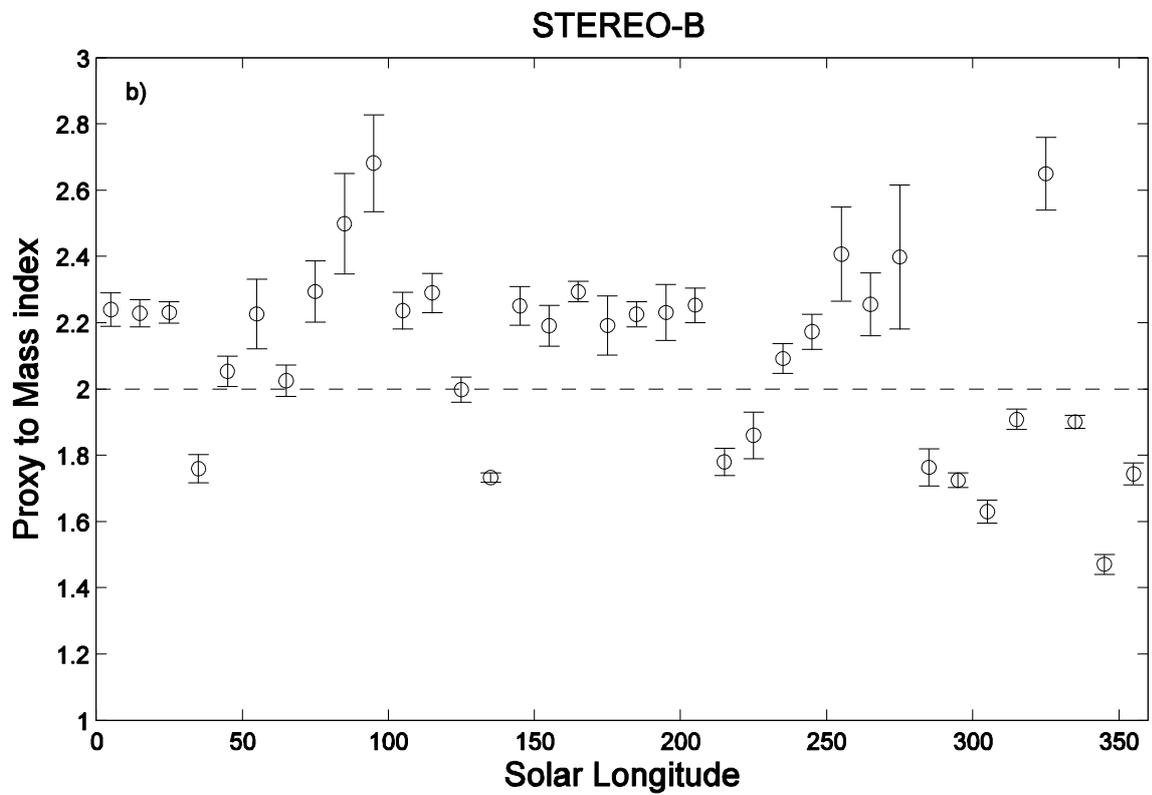

Figure 7